\RequirePackage{fix-cm}
\documentclass{svjour3}                     
\smartqed  
\usepackage{graphicx}
\begin{document}

\title{The influence of the surface roughness on dielectric function of two-dimensional electron gas
}
\subtitle{A Lindhard model approach}

\titlerunning{The influence of the surface roughness on dielectric function}        

\author{A. Phirouznia \and L. Javadian \and J. Poursamad Bonab \and K. Jamshidi-Ghaleh
}


\institute{A. Phirouznia \at
              Department of Physics, Azarbaijan Shahid Madani
University, 53714-161, Tabriz, Iran \\
               \email{Phirouznia@azaruniv.ac.ir}           
           \and
           L. Javadian  \at
           Department of Laser and
Optical Engineering University of Bonab, 5551761167 Bonab, Iran.
              \and
              J. Poursamad Bonab \at
Department of Laser and
Optical Engineering University of Bonab, 5551761167 Bonab, Iran.
\and
K. Jamshidi-Ghaleh \at
Department of Physics, Azarbaijan Shahid Madani
University, 53714-161, Tabriz, Iran
}

\date{Received: date / Accepted: date}

\maketitle

\begin{abstract}
Low field response function calculations have been
performed on a two-dimensional electron gas with well-defined
electron-surface roughness scattering. The Lindhard model was
employed to compute the response function. In particular, detailed
investigations were made on the system searching for an interplay
between surface roughness with well-defined correlation function,
(characterizes by asperity height and correlation length) spatial
confinement and the dielectric function. We analyze to what extent
the normal behavior and functionality of dielectric function of
two-dimensional devices are modified by random scattering events
caused by the contribution from the surface roughness. Results of
the current work indicate that contribution of the surface roughness
on scattering and absorption process could not be considered as an
underestimating effect. We find, however, that functionality of the
dielectric function seems to be quite independent of the particular
roughness features.
\end{abstract}

\section{Introduction}
Physics of low-dimensional systems has become an intense research
field in the last decades. Advances in material fabrication,
submicrometer technology and ultrathin film manufacturing have
opened a new field in understanding the physical processes.
Considering the remarkable
progress in empirical manufacturing of low-dimensional systems and
nano-structures and very high applications of this type of systems
in electronic and optical devices, a vast area has been developed
for physics of low-dimensional systems and nano-structure. Meanwhile
some important physical effects, such as quantum Hall effect, shows
that this research field could be very rich for fundamental studies
\cite{Girvin,Landauer,Buttiker,Zhao,He,Thomas,Starikov,Bagraev}.
\\
The low-dimensional objects can be utilized as components of the
electronic devices. These low-dimensional structures can provide new
functionalities for new generation of electronic devices. In
addition to the applicable aspects, these systems could be
considered an accurate test for quantum mechanical properties. Such
structures were first grown using molecular beam epitaxy (MBE)
technique \cite{5Smith}, meanwhile various techniques have been
recently developed for low-dimensional system fabrication.
\\
Today, non-homogeneous low-dimensional structures might be grown
from any substance at interfaces. These structures form the basis of
quantum well devices. In these structures, electron wave vector is
quantized along the confining electric field. Certain experiments
have shown that the quantization of the vertical component of wave
vector really happens \cite{6Grondin}. In an experiment, it was
discovered by measuring optical absorption in a multi-quantum well
(MQW) that absorption rate increased in specific wavelengths. These
results demonstrated that vertical wave vector is quantized and
empirical characteristics of absorption can be explained by
step-like density of states of two-dimensional electron gas.
\\
Due to the unavoidable surface roughness of any low dimensional
system, it is essential to understand the influence of boundary
scattering in physical phenomena.
\\
Formation of transverse modes, quantization of electron momentum in
the systems with reduced dimensions include some aspects of physics
in which surface roughness can effectively modify the optical
response and electronic transport of the system. Some of the effects
caused by roughness of quantum wells on electronic features have
been formerly calculated \cite{9Nag}.
\\
In the current work, this was done about electron-photon interaction
and optical features by considering successful models for the
roughness. Calculations have been performed for confined electrons
in a two-dimensional rough plane by introducing an appropriate
correlation function for the roughness.
\section{Theory and approach}
\label{sec:1} We have assumed an electronic two-dimensional system
in which the carriers have been confined in $x$-$y$ plane of an area
$L_x \times L_y=S$. This type of structures can be realized by a
semiconductor quantum well. The system has been assumed to be
subjected to an external field characterizes by a vector potential,
$\textbf{A}(\textbf{r})$. This system can be described by the
following Hamiltonian
\begin{equation}
\hat{H}=\frac{1}{2m}(\textbf{P}-\frac{e}{c}\textbf{A})^2+V_{conf}(z)+\Delta
V_{r}(\textbf{r}),
\end{equation}
where $\textbf{P}$ is the momentum operator, $V_{conf}(z)$ is the
confining electric potential along the $z$ axis, $\Delta V_{r}$ is
the potential introduced by roughness and $m$ is the electron mass.
Then the Hamiltonian of the system in the Coulomb gauge,
 $\partial_iA_i=0$, reads
\begin{equation}
\hat{H}=\frac{\textbf{P}^2}{2m}+V_{conf}(z)-\frac{e}{2mc}\textbf{A}.\textbf{P}+\frac{e^2}{2mc^2}A^2+\Delta
V_{r}(\textbf{r}),
\end{equation}
The vector potential is given as follows
\begin{eqnarray}
\textbf{A}(r,t)&=&{{A}_{0}}\hat{\textbf{e}} {{e}^{i(q.r-\omega t)}}+c.c,\nonumber\\
H_{ep}&=&\frac{e{{A}_{0}}}{mc}{{e}^{i(q.r-\omega
t)}}\hat{\textbf{e}}.\textbf{P}+\frac{e{{A}_{0}}}{mc}{{e}^{-i(q.r-\omega
t)}}\hat{\textbf{e}}.\textbf{P},
\end{eqnarray}
in which $\hat{\textbf{e}}$ is the polarization vector and ${A}_{0}$
denotes the amplitude of the vector potential associated with an
external electromagnetic field.
\\
For this two-dimensional electron gas (2DEG) system in $x$-$y$
plane, we can assume an average thickness $L_z$ with a random rough
boundary at $z=L_z$ \cite{Meyerovich}
\begin{equation}
z=0,~~~~~z=L_z+\Delta(\textbf{r}),
\end{equation}
where the roughness of the system characterizes by
$\Delta(\textbf{r})$ which denotes deviation from the perfect
two-dimensional plane at $\textbf{r}=x\hat{i}+y\hat{j}$.
\\
If we choose the simple quantum box transverse modes in $z$
direction given by $E_{n_z}=(\hbar\pi n_z)^2/(2mL_z^2)$, where the
eigenvalues of the $H_0=\textbf{P}^2/(2m)+V_{conf}(z)$ can be
written as $E_{kn}=E_{n_z}+\hbar^2k^2/(2m)$. Keeping only terms up
to linear in $\textbf{A}$ we arrive at the following Hamiltonian
\begin{eqnarray}
H&=&H_0-e/(2mc)\textbf{A}.\textbf{P}+\Delta V_{r}(\textbf{r})\\
&=&H_0+H_{ep}+\Delta V_{r}(\textbf{r}),
\end{eqnarray}
in which $H_{ep}$denotes the electron photon interaction. \\
 Then at
the limit of $\Delta(\textbf{r})/L_z\ll 1$ and for a given
transverse band, ${n_z}$,
the roughness potential may then be written as 
\begin{eqnarray}
\Delta
V_{r}(\textbf{r})=&&E_{n_z}(L_z+\Delta(\textbf{r}))-E_{n_z}(L_z)
\nonumber\\
\simeq&&E_{n_z}(L_z)\frac{2\Delta(\textbf{r})}{L_z}
\end{eqnarray}
\\
Regarding the fact that $\Delta(\textbf{r})$ has been assigned
randomly, this deviation should satisfy the requirement
$<\Delta(\textbf{r})>=0$ which results in $<\Delta
V_r(\textbf{r})>=0$ in which $<...>$ denotes the spatial average of
a typical quantity.
\\
Gaussian type correlation functions are generally employed for
roughness fluctuations. Meanwhile, the exponential correlation
functions lead to a better fit with experimental results
\cite{Meyerovich,Goodnick}. Due to this fact an exponential
correlation function, has been employed as follows
\begin{eqnarray}
\left\langle \Delta V_r(r)\Delta V_r({r}') \right\rangle
&=&{{E}_{n_z}}^{2}\frac{4}{{{L}_{z}}^{2}} \left\langle \Delta
(r)\Delta
({r}') \right\rangle \nonumber\\
&=&{{E}_{n_z}}^{2}\frac{4}{{{L}_{z}}^{2}}\Delta
_{0}^{2}{{e}^{-\left| r-{r}' \right|/\Lambda }}.
\end{eqnarray}
In the presence of the above mentioned relaxation mechanisms i.e.
the external field and the surface roughness total scattering rate
in the system is given by
\begin{eqnarray}
W(q,\omega )=&&\frac{4\pi }{\hbar }{{\sum\limits_{i,j} {\left|
\left\langle  {{\psi }_{i}}
\right|(-e/(2mc)\textbf{A}.\textbf{P}+\Delta
V_{r}(\textbf{r}))\left| {{\psi }_{j}} \right\rangle
\right|}}^{2}}\nonumber\\&& \times\delta ({{E}_{j}}-{{E}_{i}}-\hbar
\omega )\left[ f({{E}_{i}})-f({{E}_{j}}) \right],
\end{eqnarray}
where $\left| {{\psi }_{i}} \right\rangle$ is the eigen-state of the
unperturbed Hamiltonian, $H_0$, and $f({{E}})$ is the Fermi-Dirac
distribution function. \\
The matrix elements of the relaxation couplings can be easily find
to be
\begin{eqnarray}
{{\left|{{H}^{~~ {k}'{{{{n}'}}_{z}}}_{ep,k{{n}_{z}}}}+\Delta
V_{r,k{{n}_{z}}}^{{k}'{{{{n}'}}_{z}}} \right|}^{2}}=&&{{\left|
{{H}^{~~ {k}'{{{{n}'}}_{z}}}_{ep,k{{n}_{z}}}} \right|}^{2}}+{{\left|
\Delta V_{r,k{{n}_{z}}}^{{k}'{{{{n}'}}_{z}}} \right|}^{2}} \\+&&2
\mathrm{Re}(({{H}^{~~
{k}'{{{{n}'}}_{z}}}_{ep,k{{n}_{z}}}}))^*\times\Delta
V_{r,k{{n}_{z}}}^{{k}'{{{{n}'}}_{z}}}).\nonumber
\end{eqnarray}
The third term of the above expression could be neglected since
\begin{eqnarray}
\left| \left\langle  k{{n}_{z}} \right|\Delta V_r(r)\left|
{k}'{{{{n}'}}_{z}} \right\rangle  \right|\approx\left\langle  k{{n}_{z}} \right|\overline{\Delta
V_r}(\textbf{r})\left|
{k}'{{{{n}'}}_{z}} \right\rangle =0.
\end{eqnarray}
In which $\overline{\Delta
V_r}(\textbf{r})=<\Delta V_r(r)>$. Meanwhile the second term, $|\left\langle k{{n}_{z}} \right|\Delta
V_r(r)\left| {k}'{{{{n}'}}_{z}} \right\rangle|^2={{\left| \Delta
V_{r,k{{n}_{z}}}^{{k}'{{{{n}'}}_{z}}} \right|}^{2}}$ can be
approximated as
\begin{eqnarray}
{{\left| \Delta V_{r,k{{n}_{z}}}^{{k}'{{{{n}'}}_{z}}}
\right|}^{2}}=&&\frac{\delta_{n_z,n'_z}}{S}{{\int{\int{{{e}^{-i(k-k').(r-{r}')}}}\Delta
V_r(r)\Delta
V_r(r'){{d}^{2}}r{{d}^{2}}r'}}}\nonumber\\
\simeq&&\frac{\delta_{n_z,n'_z}}{S}{{\int{\int{{{e}^{-i(k-k').(r-{r}')}}}\langle\Delta
V_r(r)\Delta V_r(r')\rangle{{d}^{2}}r{{d}^{2}}r'}}}\nonumber\\
=&&\frac{\delta_{n_z,n'_z}}{S}\int{\int{{{e}^{i\left| r-{r}'
\right|\left| k-{k}' \right|\cos \theta }}}{{e}^{-\left| r-{r}'
\right|/\Lambda }}{{d}^{2}}r{{d}^{2}}{r}'}\nonumber\\
=&&\frac{4}{{{L}_{z}}^{2}}{{E}_{n_z}}^{2}\Delta _{0}^{2}\frac{2\pi
{{\Lambda }^{2}}}{{{\left( 1+{{q}^{2}}{{\Lambda }^{2}}
\right)}^{\frac{3}{2}}}}\delta_{n_z,n'_z}.
\label{rough}
\end{eqnarray}
Accordingly the transition rate can be decomposed as
\begin{eqnarray}
{W}_(q,\omega )={{W}_{1}}(q,\omega )+{{W}_{2}}(q,\omega ),
\end{eqnarray}
in which
\begin{eqnarray}
{{W}_{1}}(q,\omega )=&&\frac{2\pi }{\hbar }{{\left(
\frac{e{{A}_{0}}}{mc} \right)}^{2}}
2\sum\limits_{k{{n}_{z}},{k}'{{{{n}'}}_{z}}}{{{\left| \left\langle
k{{n}_{z}} \right|{{e}^{iq.r}}\textbf{e}.\textbf{P}\left| {k}'{n}'
\right\rangle
\right|}^{2}}}\nonumber\\
&&\times\delta ({{E}_{{{k}'}}}-{{E}_{k}}-\hbar \omega )\left[
f({{E}_{k}})-f({{E}_{{{k}'}}}) \right]
\end{eqnarray}
and
\begin{eqnarray}
{{W}_{2}}(q,\omega )=&&\frac{2\pi }{\hbar
}2\sum\limits_{k{{n}_{z}},{k}'{{{{n}'}}_{z}}}|\left\langle
k{{n}_{z}} \right|\Delta V_r(r)\left| {k}'{{{{n}'}}_{z}}
\right\rangle|^2
 \nonumber\\
&&\times\delta ({{E}_{{{k}'}}}-{{E}_{k}}-\hbar \omega )\left[
f({{E}_{k}})-f({{E}_{{{k}'}}}) \right].
\label{a3}
\end{eqnarray}
Therefore the contribution of the surface roughness in the dielectric function is determined by $W_2(q,\omega )$.
If we assume $\hat{e}=\hat{x}$ then
\begin{eqnarray}
 \left\langle  k {{n}_{z}} \right|{{e}^{iq.r}}{{P}_{x}}\left|
{k}'{{{{n}'}}_{z}} \right\rangle  ={{\hbar }}{{{k}'}_{x}}{{\delta
}_{{{n}_{z}}{{{{n}'}}_{z}}}}\delta ({k}'+q-k).
\end{eqnarray}
The real part of the conductivity is in the framework of the
Lindhard approach is then given by \cite{Grosso}
\begin{eqnarray}
{{\sigma }_{re}}(q,\omega )=\frac{{{c}^{2}}}{2V}\frac{\hbar \omega
W(q,\omega )}{{{\omega }^{2}}A_{0}^{2}}.
\label{a1}
\end{eqnarray}
The imaginary dielectric function is given by
\begin{eqnarray}
{{\epsilon }_{im}}(q,\omega )&=&\frac{4\pi }{\omega }{{\sigma
}_{1}}(q,\omega ) \\
 &=& \frac{\Gamma
S}{{{(\hbar \omega )}^{2}}}\left[ {{W}_{1}}+2\pi
{{(\frac{{{E}_{1}}{{\Delta }_{0}}\Lambda
}{{{L}_{z}}S})}^{2}}\frac{{{W}_{2}}}{{\gamma}} \right].
\label{a2}
\end{eqnarray}
In which $\Gamma ={e}^{2}/(E _{f}^{2}{\pi }^{2})$ and
${\gamma}={{e}^{2}}{{A}_{0}}^{2}/(2m{{c}^{2}})$.
These relations (Equations (\ref{a1})-(\ref{a2})) indicate that all of the mechanisms which contribute to the conductivity of the system
can contribute in the amount of the dielectric function as well.
\\
Similarly the real part of the dielectric function is given by the
Kramers-Kronig relation
\begin{eqnarray}
{{\epsilon }_{re}}(q,\omega )=1+\frac{1}{\pi }P\int_{-\infty
}^{+\infty }{\frac{{{\epsilon }_{im}}(q,{\omega }')}{{\omega
}'-\omega }}d{\omega }'.
\end{eqnarray}
\section{Result and Discussion}
As mentioned in the previous section we have employed Lindhard model
to formulate the influence of the surface roughness on the
dielectric function of a two dimensional electron gas. Results of
the current work have summarized in the following figures.
\\
As depicted in Figure \ref{fig1} by increasing the correlation
length of the surface roughness the imaginary part of the dielectric
function increases. Mean while increment of the correlation function
preserves the typical functionality of the imaginary dielectric
function.
\\
\begin{figure*}
\resizebox{0.75\textwidth}{!}{%
  \includegraphics{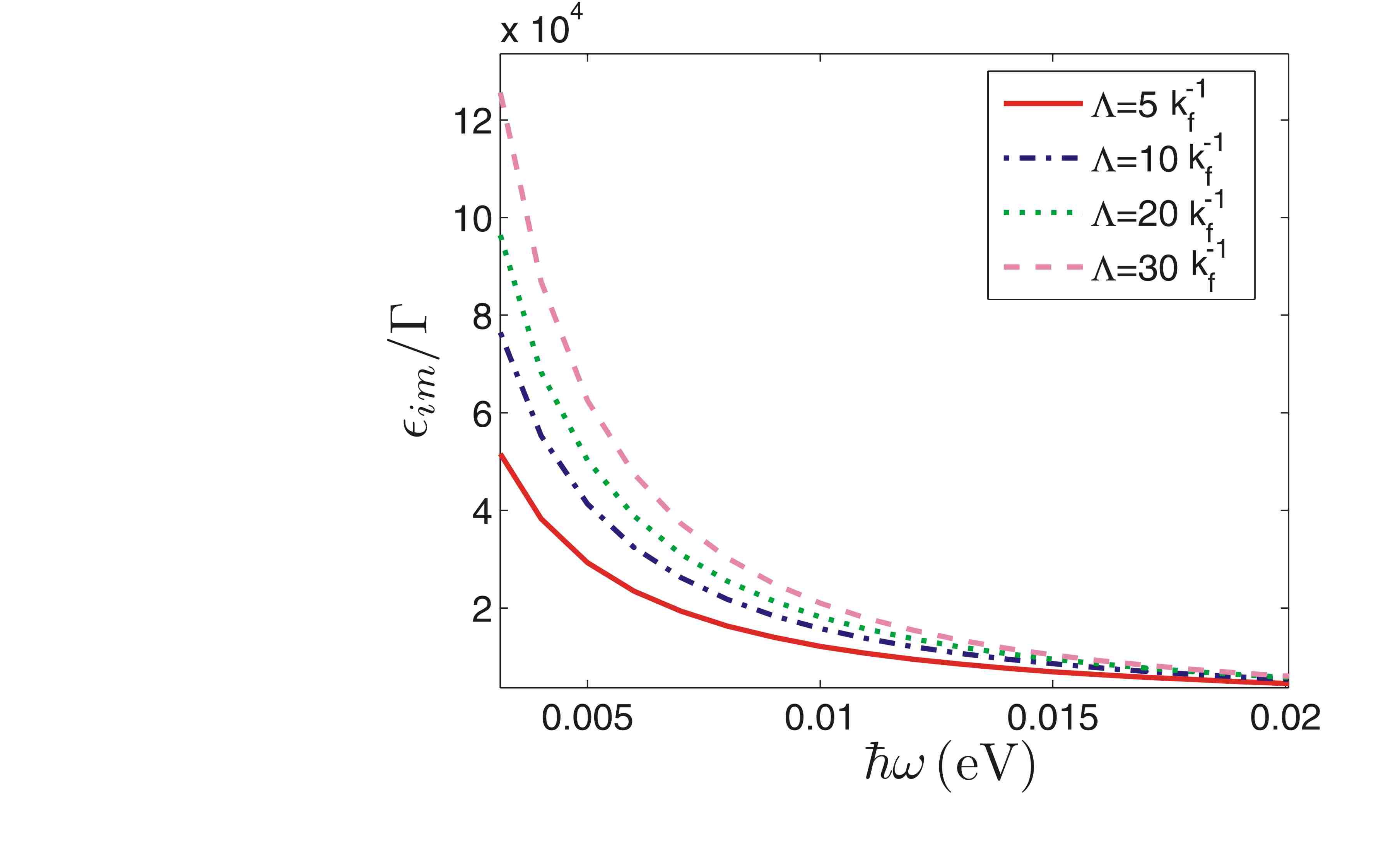}
}
\caption{Imaginary part of the dielectric function as a function of the photon energy at different correlation lengths.}
\label{fig1}       
\end{figure*}
Numerical results show a similar effect for the real part of the
dielectric function as shown in Figure \ref{fig2}. This can be
inferred by analyzing the physical meaning of the correlation
function. The effective potential of a single local roughness,
varies in spatial scale characterizes by the correlation length in the real space.
High correlation length corresponds to relatively smooth systems when $\Delta/\Lambda\ll 1$.
In this case the local rough domains have a considerable overlap.
\\
\begin{figure*}
\resizebox{0.8\textwidth}{!}{%
  \includegraphics{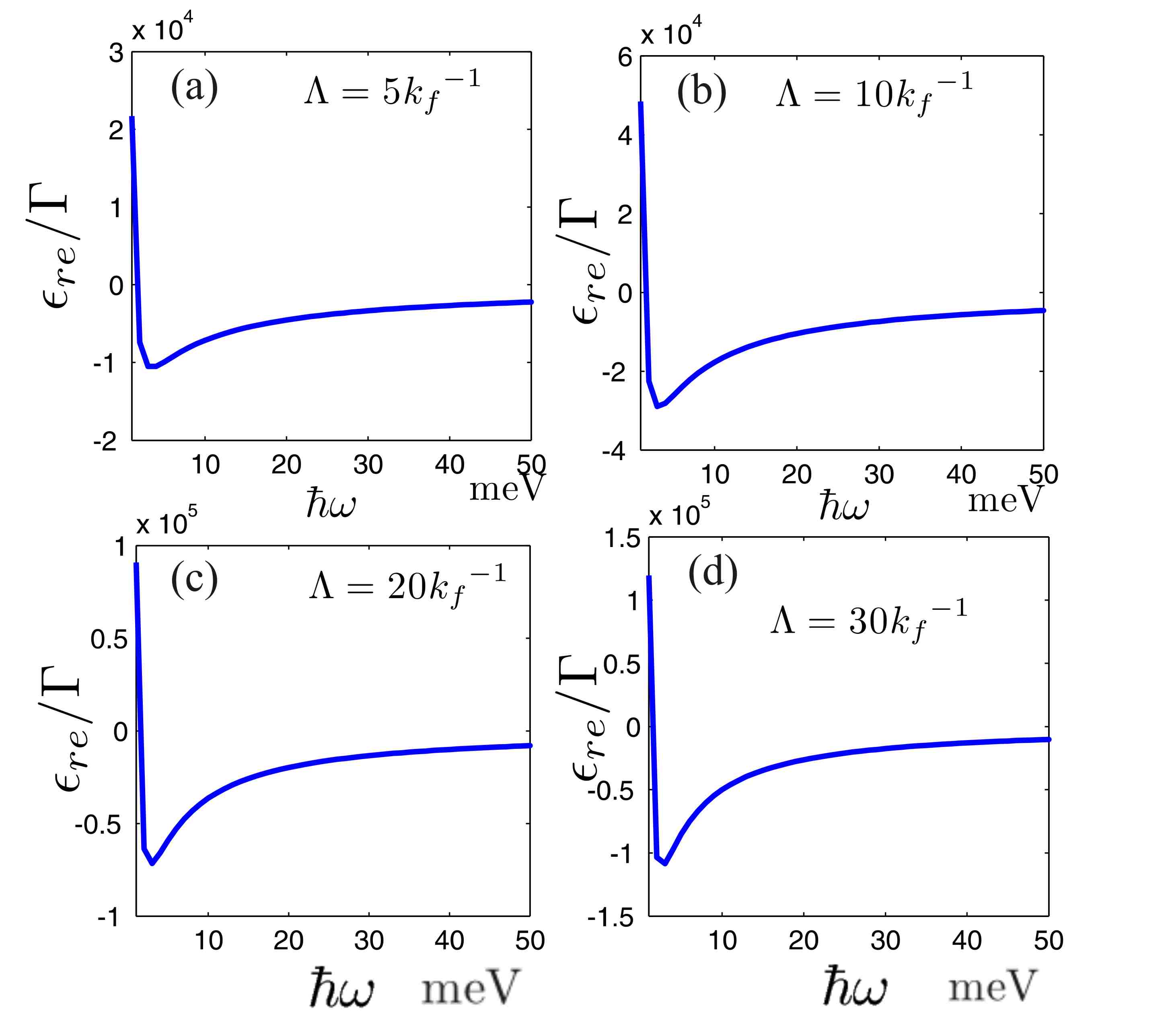}
}
\caption{Real part of the dielectric function as a function of the photon energy at different correlation lengths.}
\label{fig2}       
\end{figure*}
\begin{figure*}
\resizebox{0.75\textwidth}{!}{%
  \includegraphics{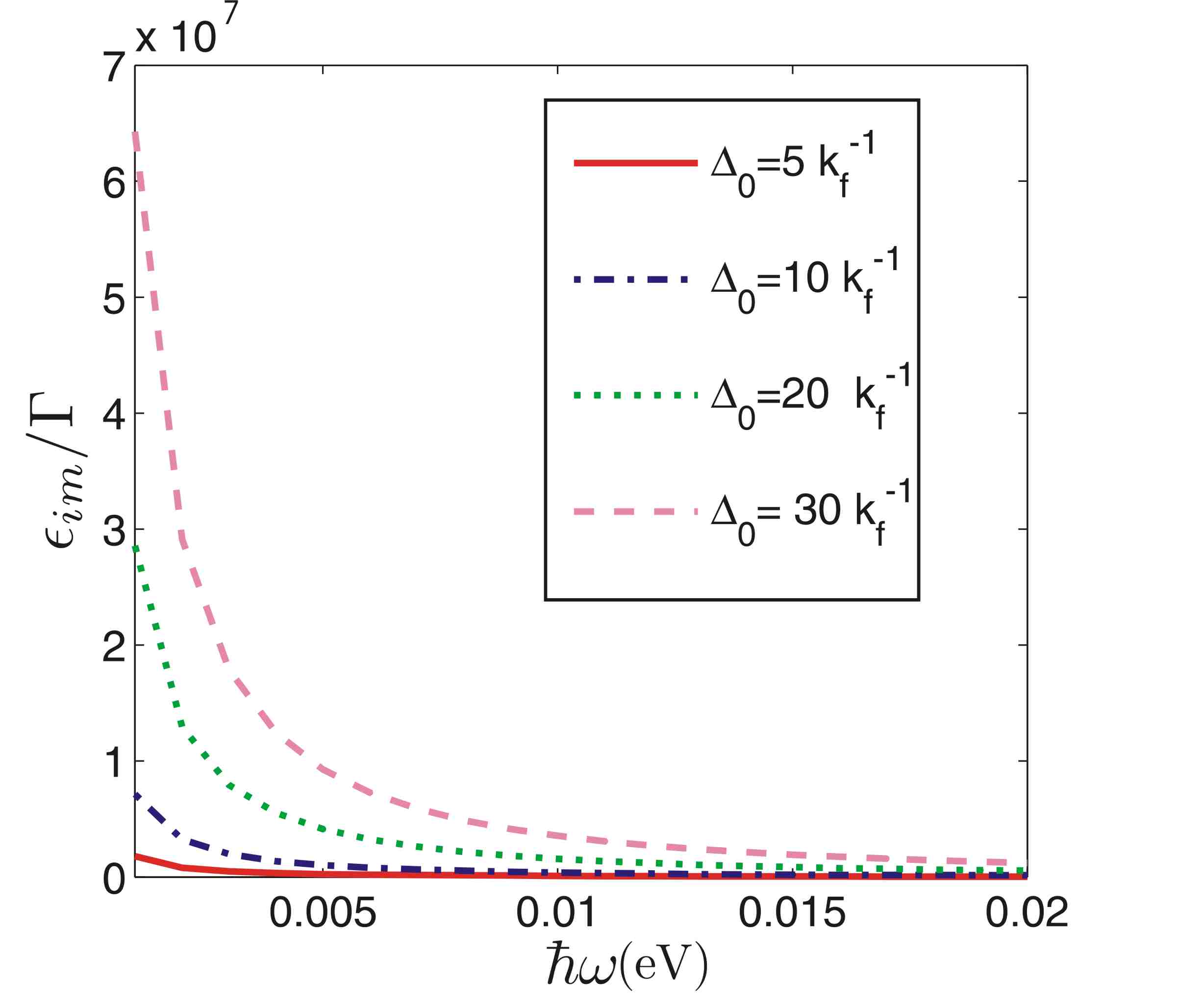}
}
\caption{Imaginary part of the dielectric function as a function of the photon energy at different asperity heights.}
\label{fig3}       
\end{figure*}
Increasing the correlation length, decreases the scattering potential
of the roughness as given in Equation \ref{rough}.
Therefore it seems that by increasing the correlation length
the imaginary part of the dielectric function (which measures the optical absorption)
should be decreased ($\lim_{\Lambda\rightarrow\infty} W_2= 0$),
however it should be noted that the $W_2$ is not a monotonic function of $\Lambda$. In fact $W_2$
can be increased by increasing the $\Lambda$ when $0<\Lambda<\sqrt{2}/q$
and decreases when $\sqrt{2}/q<\Lambda$.
\\
Meanwhile the
energy conservation rule, where enforced by the Dirac delta (Equation (\ref{a3}))
function, finally determines the transferred momwntum, $q$, and
therefore the scattering rate and the effective
contribution of the surface roughness in the absorption process. Since the
transferred momentum during a single scattering process is limited
to the range of $q< k_f$ therefore overall scattering rate increases
by increasing the correlation length of the surface
roughness. In this limit of the transferred momentum the scattering rate,
could be increased by the increasing the correlation length.
Therefore the imaginary part of the dielectric function which
measures the optical absorption increases by increasing the
correlation length.
\\
This fact will be reflected on real part of the dielectric function,
as well this fact was traced back to the Kramers, Kronig relations.
It should be noted that the functionality of the real part
dielectric function is not influenced by increasing the correlation
length however the value of the real dielectric function effectively
changes up to several order of magnitudes (Figure \ref{fig2}).
\\
Accordingly since the functionality of the real part of the
dielectric function will be preserved, therefore it seems that plasmon modes of the system
could not significantly be affected by the magnitude of the correlation length.
Meanwhile the screening length of the local charged impurities and
Friedel oscillations could be influenced by the roughness
parameters.
\\
Scattering rate of the system increases by
increasing the asperity height and this increment is independent of
the range of momentum transfer. In this case scattering rate is a
monotonic increasing function of the asperity height. Therefore as reasonably expected,
imaginary part of the dielectric function increases by increasing the asperity height (Figure \ref{fig3}).
\section{Conclusion}
In present work we have shown that the surface roughness significantly contributes
on scattering and absorption process and dielectric function. Dielectric function of the system increases by
both asperity height and correlation length of the roughness.




\end{document}